\documentclass[preprint,showpacs,preprintnumbers,amsmath,amssymb]{revtex4-2}
\usepackage{graphicx}
\usepackage{bm}
\usepackage{color}
\usepackage{xcolor}
\usepackage{caption}
\usepackage{ragged2e}
\usepackage{subcaption}
\usepackage{amssymb}
\usepackage{epsfig}
\usepackage[colorlinks=true, allcolors=blue]{hyperref}

\newcommand{\bk}[1]{\langle #1 \rangle}
\newcommand{\ket}[1]{| #1 \rangle}
\newcommand{\pdy}[1]{\frac{\partial #1}{\partial \Lambda_e}}

\newcommand{\red}[1]{\textcolor{red}{#1}}

\begin{document}

\title{Single-particle entanglement dynamics in complex systems}
\author{Devanshu Shekhar and Pragya Shukla}
\affiliation{ Department of Physics, Indian Institute of Technology, Kharagpur-721302, West Bengal, India }
\date{\today}

\widetext

\begin{abstract}

We analyze the effect of varying system conditions on the single-particle entanglement entropy for an arbitrary eigenstate of a complex system that can be described by a multiparametric Gaussian ensemble.  Our theoretical analysis leads to the identification of a single functional of the system parameters  that  governs the entropy dynamics. This reveals a  sensitivity of the entropy to collective information content, characterized by the functional, instead of the individual system details.  The functional can further be used to identify the universality classes as well as a deep web of connection underlying different quantum states.

\end{abstract}


\maketitle

\def\stackalignment{l}
\maketitle


\section{Introduction}

The standard  tools of entanglement rely on the tensor-product structure of the state space of a composite
quantum system.  While such an underlying  structure is  always present in case of distinguishable particles,  the constraints 
on the particle statistics  in case of indistinguishable ones confine the available Hilbert space, thus rendering the standard approach inapplicable.   To overcome this issue, an alternative tool has been proposed in the past: it  is based on an isomorphism (nonunique) of the Fock space to the state space of a composite (many modes) quantum system \cite{Zanardi2002}. The idea in turn motivated the concept of single particle entanglement (SPE): consider a particle confined on a bipartitioned lattice of arbitrary dimensions. The single particle   state of the lattice Hamiltonian  can then be expressed as a tensor product of  the site occupation number basis states in the second-quantized Fock space. The site entanglement in the two parts of the lattice can then be utilized as a criteria to analyze those aspects of single particle wave dynamics  where  traditional criteria such as inverse participation ratio etc. are inadequate. Some examples in this context are wave dynamics under anisotropic system conditions,  localization to delocalisation phase transition etc. Indeed, application of the entanglement statistics for characterization of the phase transitions has been studied in many previous studies \cite{psds2,psds3,Luitz2015,  Zhao2020, zhang2022,Skinner2019} but most of them have been confined to bipartite or multipartite entanglement measures. Comparatively, less attention has been paid to the single-particle entanglement analysis \cite{Pouranvari2024,Jia2008}, which, albeit counterintuitive, has been proven to be a genuine resource in quantum information processing \cite{Lee2001,Zanardi2002}. In this work, we pursue the analysis further, by developing a common mathematical formulation of single-particle entanglement entropy dynamics and apply it to two prototypical Hamiltonians, viz., the three-dimension (3D) Anderson model representing a particle moving in a random potential with its hopping confined to some nearest neighbors, and the Rosenzweig-Porter ensemble representing long range interactions in the basis space.



The first and foremost requirement for the entanglement analysis of a physical system is the knowledge of  its Hamiltonian in the physically motivated basis and thereafter its eigenstates. A  realistic system is however almost always many body, consisting of complicated interactions among its subunits. The latter  even if well-known e.g. coulomb interactions, an exact determination of the matrix representation of the Hamiltonian is often not possible; this could occur, for example, due to technical issues involved in determination of the matrix elements (e.g. calculation of the integrals either by theoretical or numerical route). The incomplete knowledge or error in their determination manifests itself by randomization of the matrix elements that can however be of different types. For example, if a matrix element can be determined upto its average value and variance, the maximum entropy hypothesis predicts its distribution to be a Gaussian. An availability of information about higher order moments can lead to non-Gaussian distributions and existence of conservation laws can result in correlated distributions or other constraints on the matrix elements e.g. column/row sum rule.  In addition, not all elements need be random,  some of them can be exactly determined. For example, for Hamiltonian of a tight binding lattice  with onsite disorder and nearest neighbour non-random hopping, the diagonals in a site basis are randomly distributed but  the off-diagonals are all non-random (given by hopping strength). Indeed, the nature and type of the distribution of the matrix elements is sensitive to various system conditions e.g. symmetry and conservation laws, dimensionality and boundary conditions, disorder etc. and can vary from one element to the other. As a consequence, the  Hamiltonian matrix is best represented  be a system-dependent random matrix, with some or all elements randomly distributed. The information about the system appear through the distribution parameters.  As expected, the randomness underlying  the matrix elements of the Hamiltonian  also manifests in its eigenstates  and the distribution of the latter can be derived, in principle, from a multivariable integration over the Hamiltonian ensemble density  (i.e. the JPDF  of the matrix elements); this is achieved  by a transformation of variables from matrix space to eigenvalue-eigenstate space and thereafter an integrating over all eigenvalues and all other eigenfunctions except one of them. The information in turn leads to statistics of the entanglement measures of the remaining eigenstate.
 
A multivariable integration of the ensemble density of the Hamiltonian  is   in general technically complicated. In addition, 
the interactions among various subunits of a many body system are  in general sensitive to a host of system conditions, both static as well dynamic type. A variation of these conditions can lead to changes in the mutual interactions among subunits and thereby the ensemble density.    This in turn is expected to manifest in the behaviour of a typical many body state and thereby entanglement measures,  This leads to serious technical issues: even if the integration is achieved by some approximations, these are often specific to system parameters and may not be applicable once the parameters are changed.  The integration route is therefore not often a viable option and  motivates a search for alternative tools which take into account arbitrary variation of the system parameters. 
 


The search for alternative tools to determine entanglement statistics motivated the complexity parameter formulation for the physical Hamiltonians represented by the multiparametric Gaussian ensembles \cite{psco, psalt, pseig}. As discussed in \cite{pseig}, the joint probability distribution function (JPDF) of the components of an arbitrary eigenstate (referred as state JPDF hereafter for brevity) for Hermitian operators represented by a multiparametric Gaussian ensembles undergoes a multiparametric evolution as the ensemble parameters vary. For a specific variation of the latter, the evolution describes a diffusion in state space and  can exactly be derived from the ensemble density.   The diffusion equation has an additional benefit: it provides a common mathematical formulation, referred as the complexity parameter formulation,  of the state JPDF for a wide range of many body systems where the system information is contained in a single parameter, i.e., the complexity parameter. We apply the above formulation in the present work and derive thereby the evolution equation for the single-particle entanglement measures and their solutions. 



We proceed as follows. We begin in section \ref{spem}  by introducing the standard definitions of the single particle entanglement entropy (SPEE). This is followed by section \ref{density} describing the single-particle models considered in this work and the respective JPDF of their matrix elements. Section \ref{diff-density} revisits  the complexity parameter formulation briefly explaining how the changing system conditions lead to an evolution of the JPDF of the matrix elements, and in turn of the eigenfunction components. They are described by a Fokker-Planck equation in terms of a parameter $Y$ which is a function of the ensemble parameters. The JPDF dynamics  of the eigenfunction components is later used to obtain the evolution equation of the single-particle entanglement in section \ref{speeEvol}. Finally, in section \ref{numerical}, we present the exact diagonalization results of the models introduced in section \ref{density} as a validation to our theoretical claims, and conclude in section \ref{conclusion} with a summary of our results and future directions.

\section{Single particle entanglement measure} \label{spem}

\begin{figure}
    \centering
    \begin{subfigure}[t]{0.45\textwidth}
        \centering
        \includegraphics[width=0.5\textwidth]{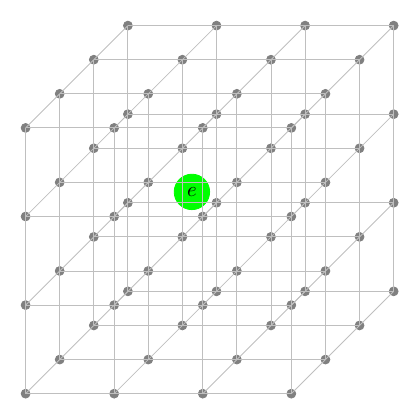}
    \end{subfigure}
    \hfill
    \begin{subfigure}[t]{0.45\textwidth}
        \centering
        \includegraphics[width=0.5\textwidth]{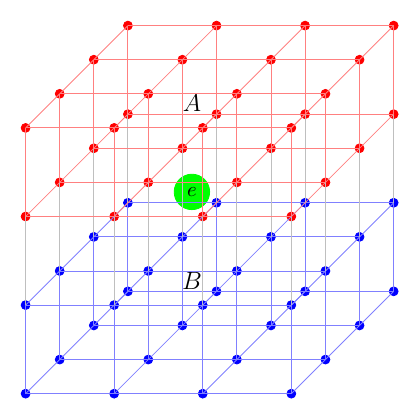}
    \end{subfigure}
    \caption{\justifying (left) An electron in a 3D lattice. (right) A horizontal bi-partitions of the lattice into two sub-parts A and B.}
    \label{lattice}
\end{figure}

Consider the dynamics of a single particle in a lattice of arbitrary dimension $D$, bipartitioned in parts  $A$ and $B$, created, e.g. by slicing the 3D lattice horizontally, vertically or in an arbitrary direction as per requirement (as illustrated in figure \ref{lattice} by a schematic diagram).
With each subpart associated with an occupation probability of the particle, it can be described by a state 
$|1\rangle$ (subpart occupied) or $|0\rangle$ (subpart empty). An arbitrary state  of the particle can  then  be written as a product over the occupation probabilities in the two parts. Based on whether the particle occupies  the sites in the subsystem A or B, two such product states  are possible. An eigenstate of the Hamiltonian $H$ can then be written as a superposition of the product states
\begin{equation}
    |\psi\rangle = |1\rangle_A |0\rangle_B + |0\rangle_A |1\rangle_B,
\label{psiae}
\end{equation}
But as the probability $P_A$ or $P_B$ for the particle to lie in the subpart $A$ or $B$ depends on the number of sites therein, this implies $|1\rangle_A = \sum_{r \in A} \psi_r |1\rangle_r \otimes_{r' \neq r} |0\rangle_{r'}$ and $|0\rangle_A = \otimes_{r \in A} |0\rangle_r$; similarly, for $B$.  The state $|\psi \rangle$ can then be written as 
\begin{equation}
    |\psi\rangle = \sum_{r \in A \cup B} \psi_r |1\rangle_r \otimes_{r' \neq r} |0\rangle_{r'},
\end{equation}
where $\ket{1}_r$, and $\ket{0}_r$ correspond to the occupation or vacancy of site $r$, respectively. 


 The formulation for the standard Von Neumann entropy of a pure quantum state is analogous to the one for the Shannon entropy. Based on similar ideas, the single particle entanglement entropy (SPEE) is defined in terms of the probability of occupation of the subparts. Let $P_A$ be the probability for the particle to occur in part $A$, 
\begin{equation}
    P_A \equiv \langle 1|1\rangle_A = \sum_{r \in A} |\psi_r|^2,
\end{equation}
the SPEE can then be defined as 
\begin{equation}
    S_A(\rho_A) = -P_A \log P_A - (1-P_A) \log (1-P_A) = -P_A \log P_A - P_B \log P_B. \label{spee}
\end{equation}
where, $P_B \equiv 1-P_A$. 
In the ergodic limit, for a balanced bi-partition implying  equal occupancy of both subparts, we have $P_A=P_B  \to 1/2$ and thereby $S_A \to \log(2)$. The other extreme limit corresponds to the particle localized in one subpart thereby implying 
$P_A=1, P_B=0$ or $P_A=0, P_B=1$;  both these possibilities correspond to $S_A \to 0$. Consequently, $S_A$ varies between 
$0$ and $\log(2)$ as the particle wavefunction starts extending from one subpart to another.



\section{Hamiltonian and ensemble density} \label{density}

Eq.(\ref{spee}) gives the SPEE for an eigenfunction of a single particle Hamiltonian. However, for cases where the Hamiltonian is best described by an ensemble, it is imperative to consider the distribution of the eigenfunction over the ensemble and therefore SPEE statistics. The latter can be derived if the theoretical formulation for the ensemble density of Hamiltonians representing the particle is available. The natural query arising then is  
how to find the appropriate as well as mathematically tractable ensemble representing a given Hamiltonian. 
The next section elucidates our approach to construct the ensembles for two prototypical single particle Hamiltonians; these are later used in the numerical analysis too.

{\bf (i) Anderson Hamiltonian:} Consider a single electron moving in  a random potential e.g. $d$-dimensional disordered lattice. Within tight-binding approximation, the dynamics can be described by the Hamiltonian \cite{psand,Biswas2000,Jia2008,Pouranvari2024}

\begin{eqnarray} 
H = \sum_{k} \epsilon_{k} |{k}\rangle \langle {k} | + \sum_{\bk{k,l}} t_{kl} |k\rangle \langle l|.
\label{ae}
\end{eqnarray}
with $t_{kl} = \langle k | H | l\rangle$ describes the tunneling amplitude of an electron between sites $k$ and $l$, and the summation $\sum_{\bk{k,l}}$ is over $z$ neighboring sites. Based on the nature of the disorder, the ensemble representing $H$ can be of various types. For example, for the independent Gaussian distributed site-energies $\epsilon_{k}$  with variance $v_{kk}=\sigma^2$, the probability density of $H_{kk}$ can be written as $\rho_{kk}(H_{kk})={\rm e}^{-(H_{kk}-\epsilon_{k})^2/2 \sigma^2}$. The hopping can be chosen to be isotropic or anisotropic, non-random or random (Gaussian). In case of non-random, anisotropic hopping $b_{kl}$ between nearest neighbour, we have $\rho_{kl}(H_{kl})=\delta\bigg(H_{kl}-b_{kl} \bigg)$ if $k,l$ form nearest neighbour pairs, and,  $\rho_{kl}(H_{kl})=\delta\bigg(H_{kl} \bigg)$ if $k,l$ are not nearest neighbours. 
The probability density $\rho(H)\equiv \prod_{\bk{k,l}}\rho_{kl}(H_{kl})$ of the ensemble can therefore  be given as
\begin{eqnarray}
\rho_1(H) = C \; \exp\left[-\sum_k \frac{\left(H_{kk}-\epsilon_{k}\right)^2}{2 \sigma^2}\right] \, \prod_{\bk{k,l}} \; \delta\left(H_{kl}-t_{kl}\right) \,  \prod_{k,l \not=n.n} \delta\left(H_{kl}\right)
\label{rho_ae1}
\end{eqnarray}
with $C$ as a normalization constant, $\prod_{\bk{k,l}}$ imply product over connected sites only (depends upon number of neighbors). Similarly, in case of random, nearest neighbor hopping, we have
\begin{eqnarray}
\rho_1(H) = C \; \exp\left[-\sum_k \frac{\left(H_{kk}-\epsilon_{k}\right)^2}{2 \sigma^2}\right] \,
 \exp\left[-\sum_{k,l=n.n.} \frac{\left(H_{kl}-t_{kl}\right)^2}{v_{kl}}\right]
\,  \prod_{k,l \not=n.n.} \delta\left(H_{kl}\right)
\label{rho_ae2}
\end{eqnarray}

Using the Gaussian limit of the delta function i.e. $\delta(x) = \red{\lim_{v \to 0}} \;  {1 \over \sqrt{2\pi v^2}}\;  {\rm e}^{-x^2/2 v^2} $, eq.(\ref{rho_ae1}) and eq.(\ref{rho_ae2}) can both be written in the form of a multi-parametric Gaussian ensemble,
%
%
\begin{eqnarray}
\rho(H; v, b) = \lim_{v \to 0} C \; \exp\left[-\sum_{k} \frac{\left(H_{kl}-\epsilon_{k}\right)^2}{2 \sigma^2} -
\sum_{k,l; k \not=l} \frac{\left(H_{kl}-t_{kl}\right)^2}{v_{kl}}\right],
\label{rho_ae}
\end{eqnarray}
with $t_{kl}=0$ for disconnected sites and the notation $\lim_{v \to 0}$ corresponds to the limit $v_{kl} \to 0$ for all $k,l$-pairs corresponding to the non-random hopping or for disconnected sites. 

{\bf (ii) Rosenzweig-Porter Ensemble:} The Rosenzweig-Porter ensemble (RPE), a prototypical model often used for spectral analysis of the localization $\to$ delocalization transition \cite{shapiro,kravtsov,buijsman}, is described by the probability density \cite{psand}
\begin{eqnarray}
\rho_2(H)  \propto 
{\rm exp}{\left[- {1 \over 2} \; \sum_{i=1}^{N} H_{ii}^2 - (1+\mu) \sum_{i,j=1; i < j}^{N} |H_{ij}|^2 \right]}, 
\label{rho_rp}
\end{eqnarray}
i.e., a Hamiltonian with normally distributed diagonal entries and the off-diagonal entries with variance controlled by an external parameter $\mu$:
\begin{equation}
    \langle \delta H_{ij}^2 \rangle = \delta_{ij} + \frac{1 - \delta_{ij}}{1 + \mu}.
    \label{var_rp}
\end{equation}
Later, in our numerics, we will set $\mu = c \, N^{\alpha}$ with $c$ and $\alpha$ as arbitrary parameters we vary.

\section{Complexity parameter space formulation of the ensemble density} \label{diff-density}

 Our prime interest in the present work is  to study the entanglement statistics of an eigenstate of a complex system represented by the Hamiltonian $H$ in single particle approximation. As the examples discussed in previous section indicate, a wide range of Hamiltonian $H$ can be described by an ensemble density
\begin{eqnarray}
\rho(H; v, b) = C \; \exp\left[-\sum_{k,l} \frac{\left(H_{kl}-b_{kl}\right)^2}{v_{kl}}\right],
\label{rho}
\end{eqnarray}
with $v_{kl}$ and $b_{kl}$ as the arbitrary variance and mean value of the matrix element $H_{kl}$. As the limit $v_{kl} \to 0$ corresponding to a non-random $H_{kl}$ taking a value $b_{kl}$, the above density can model a wide range of ensembles including sparse as well as banded ones. As obvious, the ensemble densities in eq.(\ref{rho_ae}) and eq.(\ref{rho_rp}) correspond to special cases of eq.(\ref{rho}).

%

The ensemble density in eq.(\ref{rho}) is a function of $N(N+1)/2$ matrix elements $H_{kl}$ and $N(N+1)$ ensemble parameters $v_{kl}, b_{kl}$. A variation of the matrix elements $H_{kl}$ changes the location of each matrix of the ensemble in the matrix space and thereby the moments $\langle H_{kl}^n \rangle$  averaged over the shifted ensemble. 
This give rise to many natural queries e.g. whether the new ensemble still retains its Gaussian form (requires specific conditions on $\langle H_{kl}^n \rangle$) and  can it be accessed from the initial ensemble, in the ensemble parameter space, just by varying the parameters $v_{kl}$ and $b_{kl}$.  As intuitively clear, this can only be achieved for specific dynamics in the matrix space.  Indeed, as discussed in \cite{psco}, a specific drift dynamics of the ensemble in parameter space can exactly  mimic  a diffusive dynamics with a finite drift in matrix space. In addition, the drift dynamics in the parameter space can be shown to be governed by a single function of all ensemble parameters. This is achieved by  a transformation of  the set $\{ v_{kl}, b_{kl} \}$ to another set $\{t_1, t_2, \ldots, t_M \}$  such that the multiparametric ensemble density in $v,b$ ensemble space reduces to a single parametric one, say $t_1$, while others i.e. $t_2 \ldots t_M$ appear as constants,
\begin{eqnarray}
{\partial \rho\over\partial t_1} &=&  L \, \rho;  \hspace{0.4in} {\partial \rho\over\partial t_{\alpha}} = 0, \quad \alpha >1
  \label{rhot}
\end{eqnarray}
where $L \equiv \sum_{k,l}{\partial \over 
\partial H_{kl}}\left[{g_{kl}\over 2}
  {\partial \over \partial H_{kl}} +  \gamma \, H_{kl}\; \right]$
%
%
with $g_{kl}=1$ or $2$ for $k =l$ and $k \not=l$, respectively, $\gamma$ is an arbitrary parameter related to the variance of the ensemble in the stationary limit, and 
\begin{eqnarray}  
t_1 = -{1\over  N(N+1) \gamma}  \; \; {\rm ln}\left[ \prod_{k \le  l}
 |g_{kl}-2 \gamma v_{kl}| \, |b_{kl}|^2 \right] + \text{const}.
 \label{yparam}
\end{eqnarray}
To distinguish $t_1$ from $t_2, \ldots, t_M$, hereafter we replace $t_1$ by $Y$. As clear from the above, $Y$ turns out to be an average distribution parameter, a measure of average uncertainty of system, also referred to as the ensemble \textit{complexity parameter}.


\section{Evolution of the Single-particle entanglement with system parameters} \label{speeEvol}

An eigenfunction $\Psi$ of the Hamiltonian $H$ can in principle be obtained by solving the eigenvalue equation $H \, \Psi = e \, \Psi$.    Let $z_1, \cdots, z_N$  denote the $N$  components of the state $\Psi$ in the  basis in which $H$ is represented by the ensemble density in eq.(\ref{rho}), with $\mathcal{P}_{\psi}(z_1, \cdots, z_n)$ as the JPDF of the components. As the single particle entanglement measures depend upon the eigenfunction statistics in the site basis, it suffices to start from the evolution equation of the JPDF of eigenfunction components in that basis. As discussed in detail in  \cite{pseig},  the complexity parameter based evolution of $\mathcal{P}_{\psi}$ can be given as 


\begin{equation}
{1\over \chi} \, \pdy{\mathcal{P}_{\psi}} = \sum_{m,n = 1}^N \frac{\partial^2}{\partial z_n \partial z_m} h_2 + \sum_{n=1}^N \frac{\partial }{\partial z_n} h_1,
    \label{dpn2}
\end{equation}
where, $h_1 \equiv (N-1)  \, z_n \, \mathcal{P}_{\psi}$, and $h_2 \equiv  (\delta_{mn} - z_n z_m) \mathcal{P}_{\psi}$ and 

\begin{eqnarray}
   \Lambda_e(e) = (\beta^2/4) \, (Y - Y_0) \, R_{local}^2. 
   \label{Lambda}
\end{eqnarray}
Here $ R_{local}(e) = \frac{1}{\Delta_{local}} = \frac{\xi ^d}{\Delta_e \, N}$ refers to the local mean level density at the energy $e$, $\xi$ is the ensemble averaged  localization length of the eigenfunction $\psi$, $d$ is the physical dimension of the system, and $\Delta_e$ iss the mean level spacing \cite{pseig,tmchiral}. The length $\xi$ enters in the formulation though its relation with typical intensity of the eigenfunction: $ |z_{nk}|^2 \sim  \xi_k^{-d}$.  Here 
\begin{eqnarray}
 \chi  &=& 1  \hspace{0.3in} {\rm for} \;\;  \mu < \xi_k^d,  \nonumber \\
 &\sim & {\mu \over \xi_k^d} \hspace{0.3in} {\rm for}  \;\; \mu > \xi_k^d
   \label{chi}
\end{eqnarray}
with $\mu=[{\rm e}^{2\eta (Y-Y_0)}-1]^{-1}$ and $\eta={\rm e}^{-\gamma (Y-Y_0)}$. 




\subsection{Dynamics of SPEE average}

Writing components $\psi_n$ of $\psi$  as $z_n$ (for notational simplification as well as to retain same notations as used in \cite{pseig}), we now have $P_A = \sum_{n \in A} z_n^2$ and $P_B = \sum_{n \in B} z_n^2$.  With $S_A$ defined in eq. \eqref{spee}, its ensemble average  can now be obtained by  averaging over the ensemble density $P_{\psi}$ of the eigenfunction $\psi$,
\begin{equation}
  {\bk{S_A}} = \int S_A \; {P_{\psi}} \, Dz,
\label{sp0}    
\end{equation}
where, $Dz = \prod_{i=1}^N dz_i$.  From the above definition, a complexity parameter dependence  of 
${P_{\psi}}$ also implies the same for $ {\bk{S_A}}$ and its evolution  can be derived from eq.(\ref{dpn2}). 
The steps are  as follows. From  eq. \eqref{dpn2}, we have 
\begin{equation}
    \pdy{\bk{S_A}} = \int S_A \pdy{P_{\psi}} \, Dz,
\label{sp1}    
\end{equation}
 A substitution of  eq.(\ref{dpn2}) in the right side of the above equation leads to $ \pdy{\bk{S_A}} =I_1 +I_2$ 
where 

\begin{eqnarray}
I_1 &=&  \sum_n \int S_A \, \left(\frac{\partial }{\partial z_n} h_1\right) \, \mathcal{P}_{\psi} \, Dz, \\
I_2 &=& \sum_{m,n} \int S_A \, \left(\frac{\partial^2}{\partial z_n \partial z_m} h_2\right) \, \mathcal{P}_{\psi} \, Dz. 
\end{eqnarray}


%

The above integrals can be solved by repeated partial integrations but this requires the terms 
dependent on as to how $S_A$ responds to a small change in the eigenfunction components.
Here we give the required rates of changes for the analysis,
\begin{equation}
    \frac{\partial S_A}{\partial z_n}=
    \begin{cases}
    -2 z_n \log P_A - 2 z_n, \; n \in A\\
    -2 z_n \log P_B - 2 z_n, \; n \in B
    \end{cases}
    \label{dsa1}
\end{equation}
and
\begin{equation}
    \frac{\partial^2 S_A}{\partial z_m  \partial z_n}=
    \begin{cases}
    -2 (1+\log P_A)\delta _{mn} - \frac{4z_n z_m}{P_A}, \; m,n \in A\\
    -2 (1+\log P_B)\delta _{mn} - \frac{4z_n z_m}{P_B}, \; m,n \in B
  \end{cases}
    \label{dsa2}
\end{equation}


 With repeated  partial integration of $I_1$ and $I_2$ 
 and subsequent use of the relations in eqs. \eqref{dsa1} and \eqref{dsa2}, 
 eq.(\ref{sp1})  can be further simplified.  As our focus here is in In large $N$ limit, 
 it can be rewritten as 
 


\begin{equation}
 {1\over 2 N \chi}   \pdy{\bk{S_A}}  +  \bk{S_A} \approx  - c_A  \; \bk{\log P_A} - c_B \; \bk{\log P_B}.
\label{sp3}        
\end{equation}
with $c_A= {N_A \over N}$ and $c_B= {N_B \over N}$.
A general solution of the above equation can now be given as 
\begin{equation}
 \bk{S_A}  = c_1  \; {\rm e}^{-2 N \chi \Lambda}  \ -  \; {\rm e}^{-2 N \chi \Lambda}  \int_0^{2 N \chi \Lambda} \bigg(c_A \; \bk{\log P_A} +  c_B \; \bk{\log P_B} \bigg) \; {\rm e}^{x} \;  {\rm d} x.
\label{spp3}        
\end{equation}
with $c_1$ as the constant of integration to be determined from the initial condition. Noting that either $P_A=1, P_B=0$ or $P_A=0, P_B=1$ in the separability limit, the term $\log(P_A P_B)$ in the integral in eq.(\ref{spp3})  diverges at the lower integration limit $\Lambda=0$  if the  initial state is separable. 
Choosing a weak   separability limit, with $P_A=1- \varepsilon, P_B=\varepsilon$ or $P_A=\varepsilon, P_B=1-\varepsilon$ with $0 < \varepsilon \ll 1$,  as the initial condition however gives $S_A \approx  \varepsilon (1-\varepsilon - \log  \varepsilon)$ for $\Lambda=0$.  Eq.(\ref{spp3}) then implies $c_1 \approx 
 \varepsilon (1-\varepsilon - \log  \varepsilon )$. Using $\lim_{\varepsilon  \to 0}  \varepsilon \log  \varepsilon  \to 0$ , this gives $c_1=0$  in the limit $\varepsilon \to 0$.





As clear from eq.(\ref{sp3}), the determination  of $\Lambda$-dependence of $\bk{S_A}$ requires a  prior knowledge of the $\Lambda$-dependence  of $\bk{\log P_A P_B}$. But  this in turn depends on the  negative moments of $P_A$ and leads to a set of hierarchical equations with no available solution; (this can be seen by following the similar steps used  for the derivation of the moments of  $\log(P_A P_B)$).  To overcome the technical difficult, the only option available at this stage is to consider a crude 
approximation i.e.  replace $\bk{\log P_A} \approx \log\bk{P_A}$ and similarly for $P_B$  (known to be valid for annealed randomness) in eq.(\ref{sp3}).  The $\Lambda$-dependence of  $\bk{P_A}$ can now be determined proceeding as in  $\bk{S_A}$-case. By definition, we have
\begin{eqnarray}
    \bk{P_A} = \int P_A \, \mathcal{P}_{\psi} \, Dz, \\
    \Rightarrow \pdy{\bk{P_A}} = \int P_A \pdy{\mathcal{P}_{\psi}} Dz,
\end{eqnarray}
where, $Dz = \prod_{i=1}^N dz_i$. Substitution of Eq. \eqref{dpn2} in the above equation, gives
\begin{equation}
{1\over \chi}    \pdy{\bk{P_A}} = \sum_{m,n} \int P_A \, \left(\frac{\partial^2 \, h_2}{\partial z_n \partial z_m} \right) \, \mathcal{P}_{\psi} \, D_z + \sum_n \int P_A \, \left(\frac{\partial  \, h_1}{\partial z_n} \right) \, \mathcal{P}_{\psi} \, D_z.
\end{equation}
Further, by repeated partial integration and realizing that the non-zero terms after differentiation in $\sum_{m,n}$ are such that $m,n \in A$, we get,
\begin{equation}
{1\over \chi}     \pdy{\bk{P_A}} = 2(N_A - N\bk{P_A}).
    \label{evolPA}
\end{equation}
As a check, we note that,  for the balanced bi-partition case $N = 2N_A$, the above equation gives expected limiting behavior $\bk{P_A} \to 1/2$ as $\Lambda \to \infty$.
%
%
Assuming the separability limit as the initial condition, the solution of eq. \eqref{evolPA} can be given as 
\begin{equation}
    \bk{P_A(\Lambda_e)} = \frac{1}{2}\left[1 + e^{-4N_A \chi \Lambda_e}\right].
    \label{avgPA}
\end{equation}
With $P_B = 1- P_A$, the above equation gives  $\bk{P_B(\Lambda_e)} = \frac{1}{2}\left[1 - e^{-4N_A \chi \Lambda_e}\right]$ and  $\bk{P_A} \sim 1$ and $1/2$  for $\chi \Lambda_e < 1/(4 N_A )$  and $\chi \Lambda_e > 1/ (4 N_A )$ respectively, thus indicating a rapid transition as a function of $\Lambda_e$ with $\chi \Lambda_e=1/(4 N_A)$ as the point of inflection.  A smooth crossover around this point can then be seen as a function of $N_A \chi \Lambda_e$; here we recall that $\chi$, defined in eq.(\ref{chi}) also changes with $Y-Y_0$. 

Substitution of $\bk{\log (P_A )} \approx - \log 2 + e^{-4N_A \chi \Lambda_e}$  and  $\bk{\log (P_B )} \approx - \log 2 - e^{-4N_A \chi \Lambda_e}$ in eq.(\ref{sp3}) now gives 
\begin{eqnarray}
 \bk{S_A} &=&  \log 2   \; (1-{\rm e}^{-4 N_A  \chi  \Lambda_e}), 
 \label{spz4}        
\end{eqnarray}
with $\chi$ and $\mu$ defined in and below  eq.(\ref{chi}) respectively. As both $\mu$ and $\xi^d$ depend on system parameters, the state under system conditions may correspond to either ${\mu < \xi^d}$ or ${\mu > \xi^d}$ case. With $\chi$ taking different values, this in turn leads to two possible solutions. But as clear from the above, a rescaling of  $\Lambda_e$ by $\chi$ is expected to map one solution on the other.

As discussed in next section, while the numerical result for $\bk{S_A}$  has the same form as the one given in eq.(\ref{spz4}), it does not exactly match.  
We believe this discrepancy arises from the approximation 
$\bk{\log x} \approx \log \bk{x}$.  
To rule out the possibility that the discrepancy arises from the approximation 
$\bk{\log x} \approx \log \bk{x}$,  we consider following intuitive approach to improve our theoretical prediction. For small $\Lambda$ ranges,  assuming that $P_A, P_B$ are perturbed slightly from their $\Lambda=0$ values i.e. $P_A = 1-\varepsilon, P_B = \varepsilon$ with $\varepsilon \ll 1$, we have  $\log(P_A P_B)   \approx  \log\varepsilon -\varepsilon$. 
For large $\Lambda$ ranges where typically $P_A = (1/2)-\varepsilon, P_B = (1/2) + \varepsilon$, we have  $\log(P_A P_B) \approx  -2 \log 2  $. This suggests a rapid decay of $\log(P_A P_B)$  from a divergence  to a constant value with increasing $\Lambda$. Based on the above insights, we conjecture, for the balanced case ($N_A=N_B$) and for $\Lambda$ governed evolution of $\langle \log(P_A P_B) \rangle$ from  a separable  initial state,  that 
\begin{eqnarray}
\langle \log(P_A P_B) \rangle  = -(2 \log 2 + b_0 \; \Lambda^{\nu-1})
\label{papbe}
\end{eqnarray}
 with $\nu >0$. The above conjecture is  in agreement with the balanced case numerics   for  \red{$\nu=0.74$} and $b_0 =0.24$ (figure \ref{papb}, numerical details discussed in next section). Substitution of eq.(\ref{papbe}) in eq.(\ref{spp3}) gives     
 

\begin{eqnarray}
 \bk{S_A}  &=& {1\over 2} \; {\rm e}^{-4 N_A  \Lambda}  \int_0^{4 N_A  \Lambda}  (2 \log 2 + b_0 x^{\nu-1}) \; {\rm e}^{x} \;  {\rm d} x. \label{spt3} \\
 & = &   \log 2   \; (1-{\rm e}^{-4 N_A \Lambda}) + {b_0\over 2} \;  B(1,\nu) (4 N_A \, \Lambda)^{\nu}  \; _1F_1 \bigg(\nu, \nu+1, 4 N_A  \Lambda \bigg) \;  {\rm e}^{-4 N_A  \Lambda} 
 \label{spt4}     
\end{eqnarray}
with $B(a,b)$ as the Beta function and $_1F_1(a,b;x)$ as the confluent Hypergeometric function. 
As discussed in next section,  notwithstanding a good agreement between $\langle \log(P_A P_B) \rangle$ numerics  and the conjecture in eq.(\ref{papbe}),  a comparison of  eq.(\ref{spt4})  with $\bk{S_A}$-numerics does not show any significant improvement over  eq.(\ref{spz4}). 



 %
  
For $\Lambda \to \infty$, the evolution approaches the steady state with  $ \pdy{\bk{S_A}}  =0$ with $\bk{P_A}  = \bk{P_B} =1/2$.  Eq.(\ref{spz4}) then gives  $\bk{S_A}_{\infty} =\log 2$.   This is consistent with the prediction based on eq.(\ref{spee}). This also lends credence, indirectly, to the approximation $\bk{\log P_A} \approx \log\bk{P_A}$ and similarly for $P_B$  (known to be valid for annealed randomness) used in eq.(\ref{sp3}).

\subsection{Dynamics of SPEE variance}

The evolution equation for the variance can similarly be obtained. Here we give the equation for the simple case   $N_A = N/2$ (also used in our numerics),
\begin{equation}
   {1\over 4 N \chi} \, \pdy{\bk{(\delta S_A)^2}} = -  \bk{(\delta S_A)^2} - \, \text{cov}(S_A, \log(P_A P_B)),
    \label{varspee}
\end{equation}
where, $\text{cov}(S_A,  \log(P_A P_B)) \equiv \bk{ S_A \log(P_A P_B)} - \bk{S_A} \bk{\log(P_A P_B)}$. The variance of the SPEE is thus determined by the covariance of $S_A$ and $\log(P_A)$. Proceeding again as in the case of eq.(\ref{sp2}), we obtain

\begin{equation}
\bk{(\delta S_A)^2}   = c_2  \; {\rm e}^{-4 N \Lambda}  \ -  \; {\rm e}^{-4 N \Lambda}  \int_0^{4 N \Lambda} \bigg(\text{cov}(S_A, \log(P_A P_B)) \bigg) \; {\rm e}^{x} \;  {\rm d} x.
\label{varsp3}        
\end{equation}
with $c_2$ as the constant of integration to be determined from the initial condition. A choice of the separability condition as the  initial condition implies $\bk{(\delta S_A)^2}=0$ for the initial state; eq.(\ref{varsp3}) then implies $c_2=0$.

As in the case of $\langle S_A \rangle$, here again determination of ${cov}(S_A, \log(P_A P_B))$ involves negative moments and requires solving a hierarchical set of equations which is technically complicated. This motivates again to consider an intuitive approach. As mentioned above, for $P_A = 1-\varepsilon, P_B = \varepsilon$ with $\varepsilon \ll 1$ (which is the case for small $\Lambda$ ranges), we have $S_A \approx \varepsilon(1-\varepsilon - \log \varepsilon) \approx \varepsilon$ and $\log(P_A P_B)   \approx  \log\varepsilon -\varepsilon$. 
This gives $S_A^2 \sim - S_A \log(P_A P_B) $. This is also valid for the case with $P_B = 1-\varepsilon, P_A = \varepsilon$. For large $\Lambda$ ranges where typically $P_A = (1/2)-\varepsilon, P_B = (1/2) + \varepsilon$, we have $S_A \approx  \log 2 + (1-2 \varepsilon) \varepsilon - (1+2 \varepsilon) \varepsilon =  \log 2  - 4 \varepsilon^2 $ and $\log(P_A P_B) \approx  -2 \log 2  $; this again suggests $S_A^2 \sim - S_A \log(P_A P_B) $.   Based on the above insights, we conjecture that  $\text{cov}(S_A, \log(P_A P_B))(\Lambda)$  as a function of  $\Lambda$ has the same mathematical form as  ${(\delta S_A)^2}$ as a function of  $\alpha \Lambda$: 
\begin{eqnarray}
\text{cov}(S_A, \log(P_A P_B))(\Lambda) \approx {(\delta S_A)^2}(\alpha \Lambda)
\label{cov1}
\end{eqnarray}
 with $\alpha$ as a rescaling factor. As displayed in figure \ref{papb}, the conjecture is  corroborated by our numerics for the Anderson case (system details given in section \ref{numerical}) which gives 
 \begin{eqnarray}
{\bk{(\delta S_A)^2}} \approx  a \, {\rm exp}\left[- b \, \bigg(\log(c \, \Lambda)^2\bigg) \right], \label{var1}\\
\text{cov}(S_A, \log(P_A P_B)) \approx  a  \, {\rm exp}\left[-  b \bigg(\log( \alpha_c \Lambda)^2\bigg) \right]
\label{cov2}
\end{eqnarray}
with $\alpha_c=7$.  As can be checked by a direct substitution,  the above functional forms are consistent with eq.(\ref{varsp3}) too. Here, $c = 10^4$ is a rescaling of $\Lambda$ to match the mean with the numerics, and $b = 0.06$ and $0.05$ for the variance and covariance respectively. The absolute error in the determination of the normalization factor $a$ from its exact value $\sqrt{b / \pi}$ is about $0.036$ and $0.58$ for the variance and the covariance curves respectively.

\section{Numerical Analysis} \label{numerical}

The theoretical analysis described in previous section  predicts the existence of an infinite range of universality classes of  SPEE statistics, characterised by the complexity parameter $\Lambda$,  among the eigenstates  if their Hamiltonians are represented by eq.(\ref{rho}).  The robustness of our theoretical prediction makes it imperative to seek their numerical verification. For this purpose, we numerically analyze the two ensembles introduced in section \ref{density} and described by eq.(\ref{rho_ae}) and eq.(\ref{rho_rp}).

{\bf Anderson Ensemble:} We exactly diagonalise an $L \times L \times L$ three-dimensional Anderson Hamiltonian (Eq. \ref{ae}) with periodic boundary condition for $L=12$ for several diagonal disorder strength parameter $w$ and for different combination of the parameters involved in the Hamiltonian, viz., the off-diagonal hopping variance $w_1$, and the sparsity parameter $k$, such that $k=1 \, (z=6)$ imply nearest neighbor interaction and $k=2 \, (z=24)$ implying next nearest neighbour interaction. Throughout this analysis, we have kept the hopping rate $t$ fixed to $0.5$. Eq. \eqref{rho_ae} gives the 
complexity parameter for this case: $Y = -\frac{N}{2 M \gamma} \alpha + C_0$,
where   $ \alpha = \ln |1 - \gamma w^2| + (z/2) \ln [|1-2\gamma w_1^2||t+\delta_{t,0}|^2]$, 
and, $M=\frac{N}{2}[N+z(1-\delta_{t,0})+1]$. Here, we consider an isotropic hopping with mean $t$ and variance $w_1$. We choose the initial condition to be the localized phase, such that $w = w_m \gg 1$. Thus, we have,
\begin{equation}
   Y-Y_0 = -\frac{N}{2 M \gamma} (\alpha - \alpha_0) \label{yae}
\end{equation}
with $\alpha_0$ as the $\alpha$ value corresponding to initial state and $\gamma$  related to the variance of the matrix elements in the ergodic limit. More specifically, in our numerical calculations we choose $\gamma \approx 0.5$, where $w_0$ is the smallest value of $w$ considered numerically. The $\Lambda$ in this case is calculated from the Eq. \eqref{Lambda} with $\xi ^d \approx \bk{I_2}^{-1}$, where $I_2$ is the inverse participation ratio (IPR) \cite{tmchiral}.

For each set of parameters, we consider $M=1000$ disorder realizations and use the standard \textit{shift-invert} diagonalization technique to calculate $1\%$ of the total eigenfunctions in the range $E \pm \delta E$ at the energy $E=0$ \cite{pietracaprina2018shift,slepc}. This in turn gives total $ M N/100$ eigenfunctions for an average analysis of SPEE for a given set of system conditions; clearly the averaging here includes both the ensemble as well as spectral averaging.
An important point worth emphasizing here is regarding the choice of spectral range $\delta E$: based on our theoretical predictions, the complexity parameter governs the SPEE statistics and its quantitative analogy is necessary to observe the analogy of the statistics for different parametric combinations. But $\Lambda$ is energy dependent too which makes it necessary in principle to consider only ensemble averaging.  To improve the statistics however we need to consider a spectral averaging over a small optimized range  such that  $\Lambda$ does not vary significantly over the range.

\begin{figure}[h]
    \centering
    
 \vspace{-0.25in}

 \includegraphics[width=0.7\textwidth]{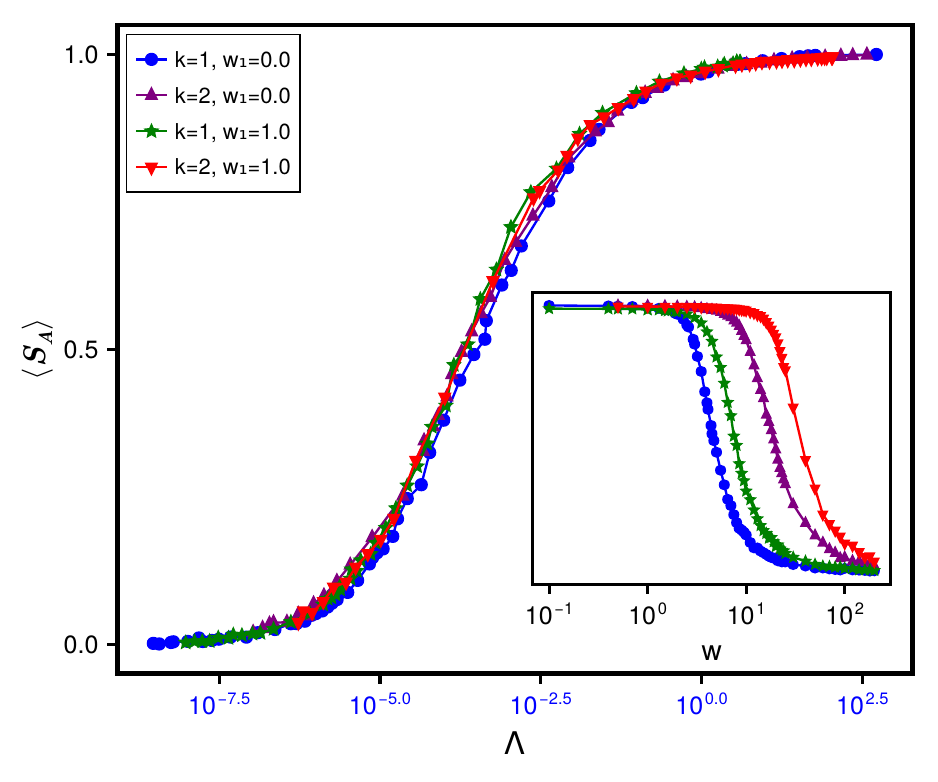}
 \includegraphics[width=0.7\textwidth]{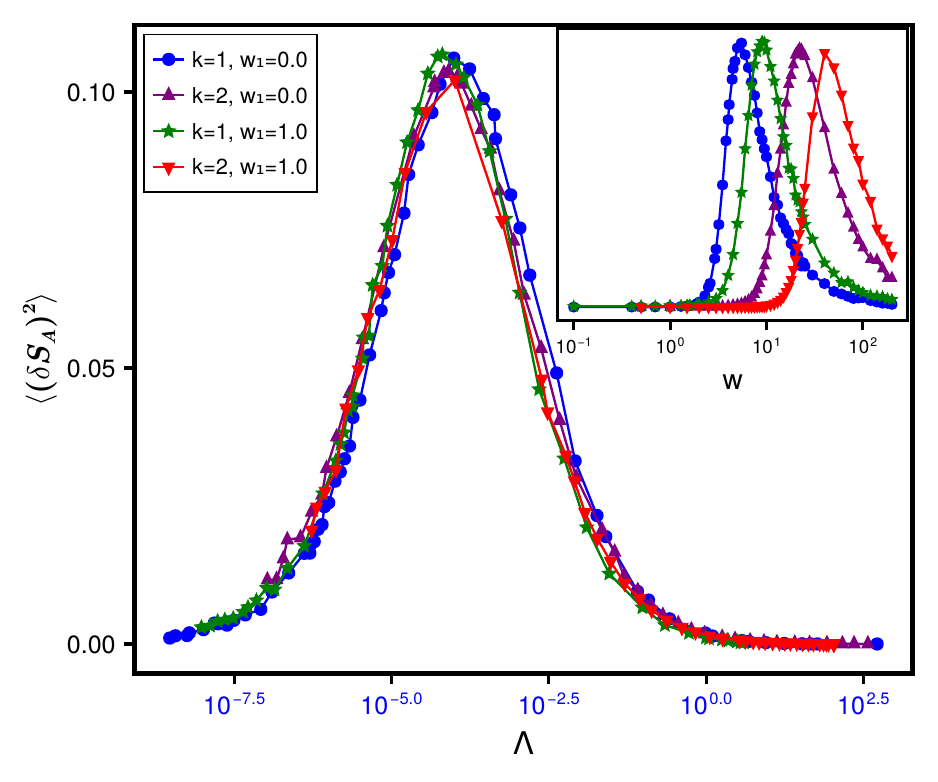}
 
 \vspace{-0.25in}
 
   \caption{\justifying \textbf{Role of $\chi$ in $\Lambda$-governed evolution:}  To illustrates the effect of ignoring the parameter $\chi$ in eq.(\ref{Lambda}) for $\Lambda$, the figure displays the $\Lambda_e$ governed evolution of average $\bk{S_A}$ and variance $\bk{\delta S_A^2}$ (in $\log_2$ base), for a cubic Anderson lattice of linear size $L=12$ is shown for  different combinations  of system parameters (for both random and non-random hopping $w_1$ and nearest neighbors ($k=1$) as well as next nearest neighbors ($k=2$) while keeping $t=0.5$ fixed). The evolution of the SPEE measures  for  $k=2$ is now visually shifted from $k=1$ but as clear from the figure \ref{rr1_ae}, the two curves can be made to collapse onto a single curve by a rescaling. For comparison, the inset also displays the  evolutions  in terms of diagonal disorder $w$.}    
  \label{r1_ae}
\end{figure}

For calculating the SPEE measures, we consider a horizontal bi-partition of the lattice through the center (figure \ref{lattice}), such that $N_A = N_B = \frac{N}{2}$. To see the effect of the type of averaging, we analyze the results  by both ways, i.e., (i) ensemble averaging only, and, (ii)  ensemble-spectral averaging (over various disorder realizations and over about $1\%$ of the total eigenstates per realization at $E=0$) for a fixed set of systems parameter.  The above analysis is repeated  by varying the parameters which gives us the average and variance of SPEE  for many $\Lambda$ values. Our analysis indicates no significant difference of the results obtained by two types of averaging. This is however  due to the choice of the eigenfunction for the analysis; we have analyzed the eigenfunction at $E=0$. The local density for Anderson ensemble in this energy range does not vary significantly, and, as a result, $\Lambda_e$ does not vary significantly within a small energy range around $E=0$.  To avoid repetition of almost same figures, here we include the figures corresponding to ensemble averaging only. The results are displayed  top and bottom panels  in figure \ref{r1_ae}.

\begin{figure}[h]
    \centering

    \vspace{-0.3in}

    \includegraphics[width=0.7\textwidth]{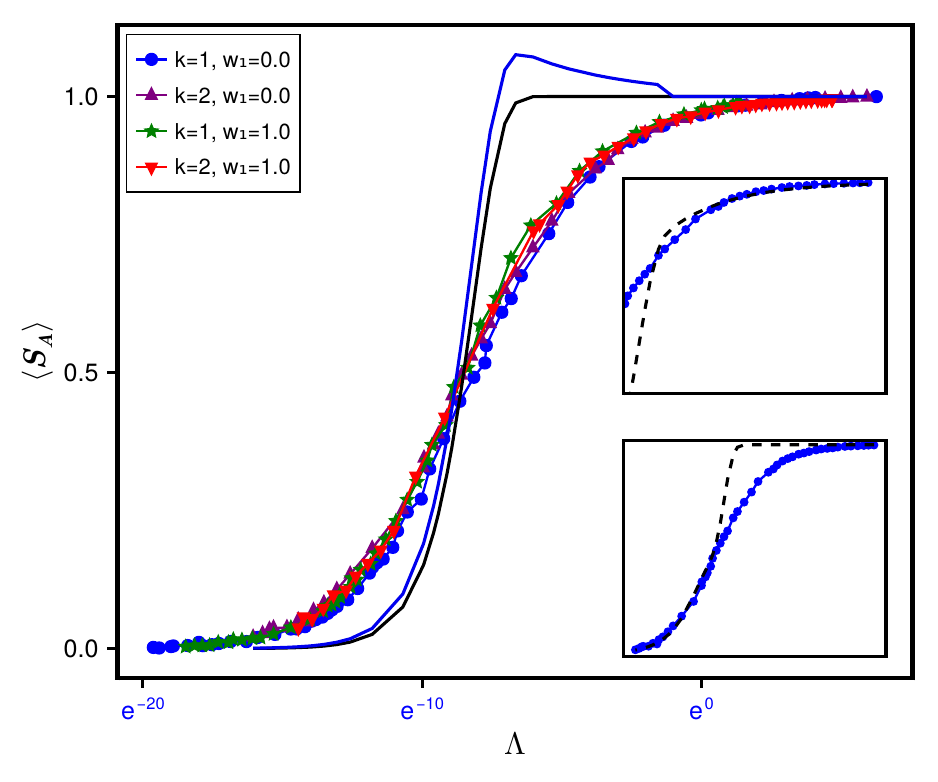}
    \includegraphics[width=0.7\textwidth]{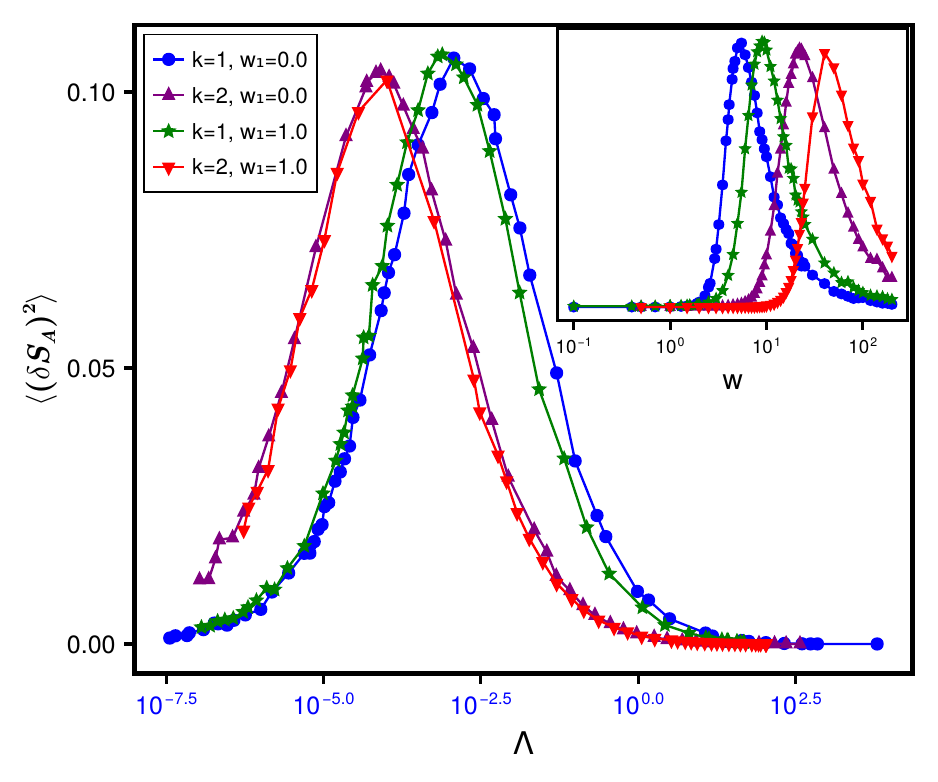}
 
    \vspace{-0.2in}
 
   \caption{\justifying \textbf{Evolution of  SPEE measures for Anderson Lattice:} 
While the other details here are same as in figure \ref{r1_ae}, the evolution parameter is  now $\Lambda$.    
The convergence of all curves to single curve in the main panel and its lack in the  insets of figure \ref{r1_ae} reveals the role of $\Lambda$ as the primary evolution parameter.  The black and blue solid lines in the $\bk{S_A}$-plot depict our theoretical predictions eq.(\ref{spz4}) and eq.(\ref{spt3}) respectively. 
%
The insets in top panel  displays a comparison with two fitted functions, namely,  $\bk{S_A} \approx (1-{\rm e}^{-2 N \Lambda})+  (2 N \Lambda)^{1/2} {\rm e}^{-8 N \Lambda}$ (fitting well in small $\Lambda$ range) and  $\bk{S_A} \approx (1-{\rm e}^{-2 N \Lambda})-0.36 (2 N \Lambda)^{-0.26}$ (fitting well for large $\Lambda$ range). The inset in bottom panel shows a comparison of numerics with eq.(\ref{var1}). }
%
%
\label{rr1_ae}
\end{figure}

To illustrate the relevance of  $\Lambda$ as a single parameter governing the SPEE dynamics, it is important to first understand as to what happens if  the dynamics  is studied in terms of one of the system parameters.
The insets in top and bottom panels of figure \ref{r1_ae}  show the SPEE variation with respect to diagonal disorder $w$ for different sets of other system parameters. As can be seen from both panels of the figure, the curves corresponding to lattice with nearest neighbour interaction ($z=6$, denoted by $k=1$) but with different hopping types, random  (off-diagonal variances $w_1=1$) as well as non random ($w_1=0$), collapse onto a single curve in terms of $\Lambda$.  The curves however collapse to a different curve if the next nearest neighbour interaction ($z=24$, denoted by $k=2$) is switched on, although the hopping strengths  remain the same ($w_1=1$ and $w_1=0$).  This is consistent with $\Lambda$ based formulation and can be explained as follows. As the ensemble averaged localization length $\xi$ and thereby $R_{local}$ varies significantly with the  nearest neighbor and next neighbor hopping, $\chi$ for the two cases is different too which in turn leads to two different curves. But, as  mentioned below eq.(\ref{spz4}) and also obvious from the shape of two curves, the two curves differ only by a rescaling. Indeed, as  can be seen from the figure \ref{rr1_ae}, a rescaling of $\Lambda$ leads to convergence of the curves for $k=1$ to those for $k=2$ thereby confirming a single-parametric evolution in terms of $\chi \, \Lambda$.

\begin{figure}[h]
    \centering
  \includegraphics[width=0.7\textwidth]{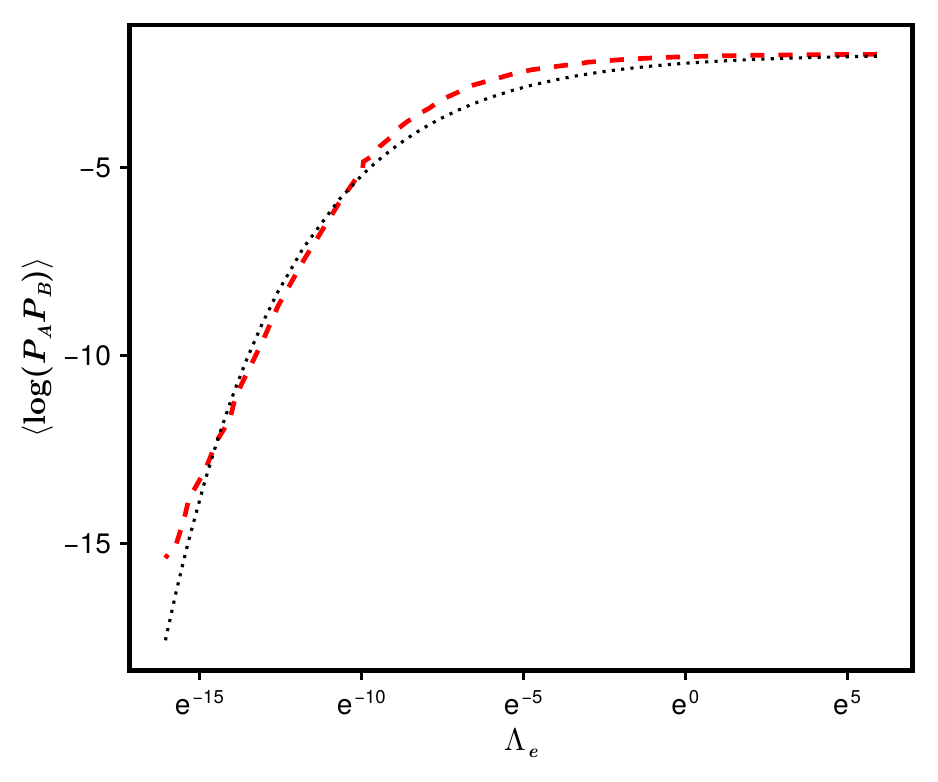}
  \includegraphics[width=0.7\textwidth]{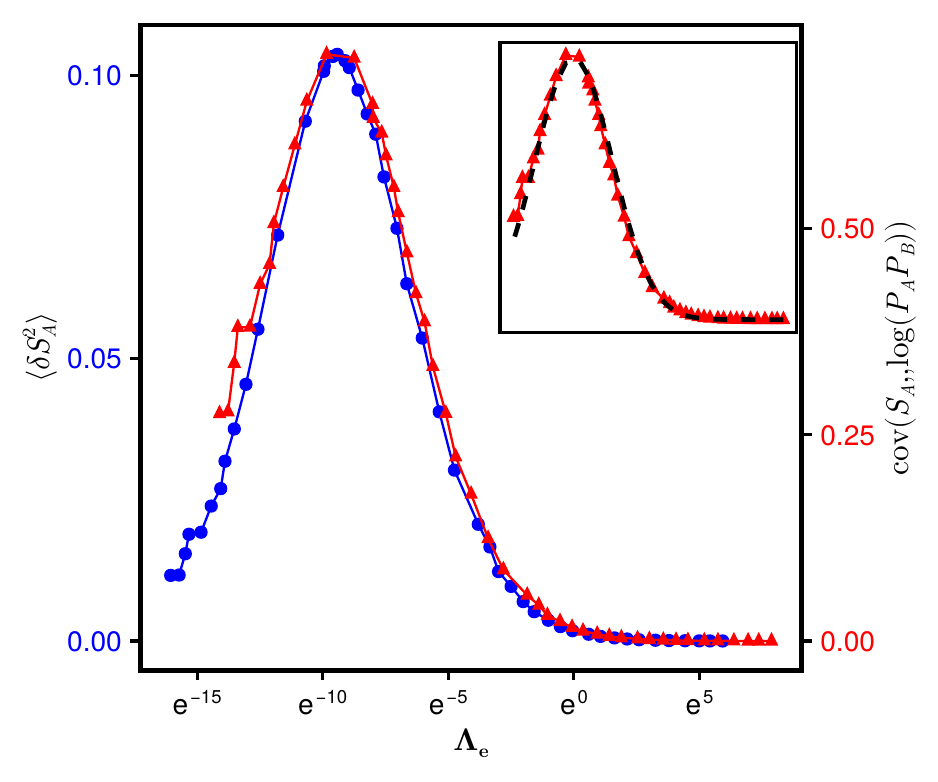}
\caption{\justifying \textbf{$\Lambda$-dependence of $\log(P_A P_B)$ :}  The figure displays the $\Lambda_e$ governed evolution of  $\bk{\log(P_A P_B)}$ and covariance $cov(S_A, \log(P_A P_B))$ (in base $\log_2$), for a cubic Anderson lattice of linear size $L=12$ for different combinations  of system parameters. The numerical results confirm  our theoretical conjectures, namely, eq.(\ref{papbe}) and eq.(\ref{cov1}). The inset in bottom panel displays a good agreement of eq.(\ref{cov2}) with numerics. }    
\label{papb}
\end{figure}

In previous section, we predicted eq.(\ref{spz4}) as a theoretical formulation for $\bk{S_A}$ (assuming 
$\bk{\log P_A} \approx \log \bk{P_A} $). Figure \ref{r1_ae} displays a comparison with eq.(\ref{spz4}); the latter  seems to predict an almost similar functional form for $\bk{S_A}$.  To improve the prediction, we  conjectured eq.(\ref{papbe}) for $\bk{\log(P_A P_B)}$ for the case $N_A=N_B$ which in turn led to eq.(\ref{spt3}) for $\bk{S_A}$. A comparison with $\bk{\log(P_A P_B)}$ numerics displayed in figure \ref{papb} shows a good agreement with eq.(\ref{papbe}) for $\nu=0.26$ and $b_0=0.24$.  This motivates us to compare  the $\bk{S_A}$-numerics with  eq.(\ref{spt3}) using same $\nu$ and $b_0$. As displayed in  figure \ref{r1_ae}, however, eq.(\ref{spt3}) does not lead to any significant improvement over eq.(\ref{spz4}). We believe the deviation could be  an artifact of computations on logarithmic scale as well finite system sizes while the theoretical predictions are derived in large $N$-limit. We also considered two fitted functions for large and small $\Lambda$; these results are also displayed in the inset of figure \ref{r1_ae}.

{\bf RP Ensemble:} We next consider the Rosenzweig-Porter Hamiltonian with its ensemble density described by Eq. \eqref{rho_rp}, and taking the initial condition as $\mu \to \infty$. We  consider the RPE states, at $E=0$, and calculate the SPEE for different  free parameters $c$ and $\alpha$ for the parameterization $\mu = c \, N^{\alpha}$, where $N$ is the size of the matrix, in Eq. \eqref{rho_rp}. Eq. \eqref{yparam} gives in this case
\begin{equation}
    Y-Y_0 = -\frac{N-1}{2(N+1)\gamma} \ln \bigg | 1-\frac{2\gamma}{1 + \mu}\bigg | \approx -\frac{1}{2\gamma} \ln \bigg | 1-\frac{2\gamma}{1 + \mu} \bigg |.
\end{equation}
For this case,  we set $\gamma = 1/2$.

\begin{figure}[h]
    \centering
 \includegraphics[width=0.7\textwidth]{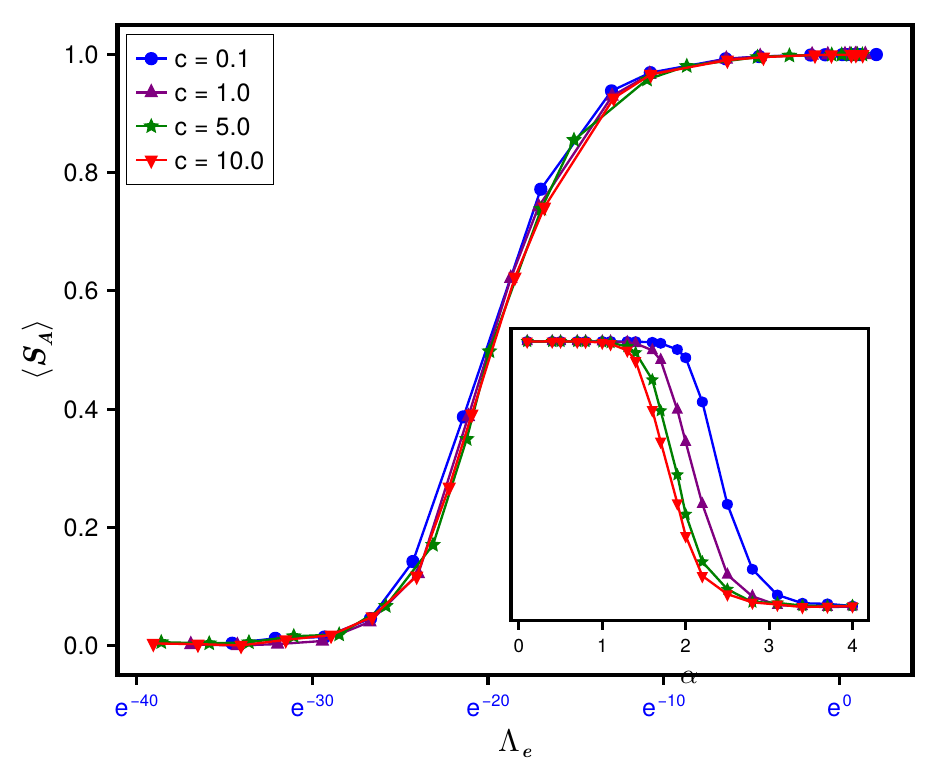}
 \includegraphics[width=0.7\textwidth]{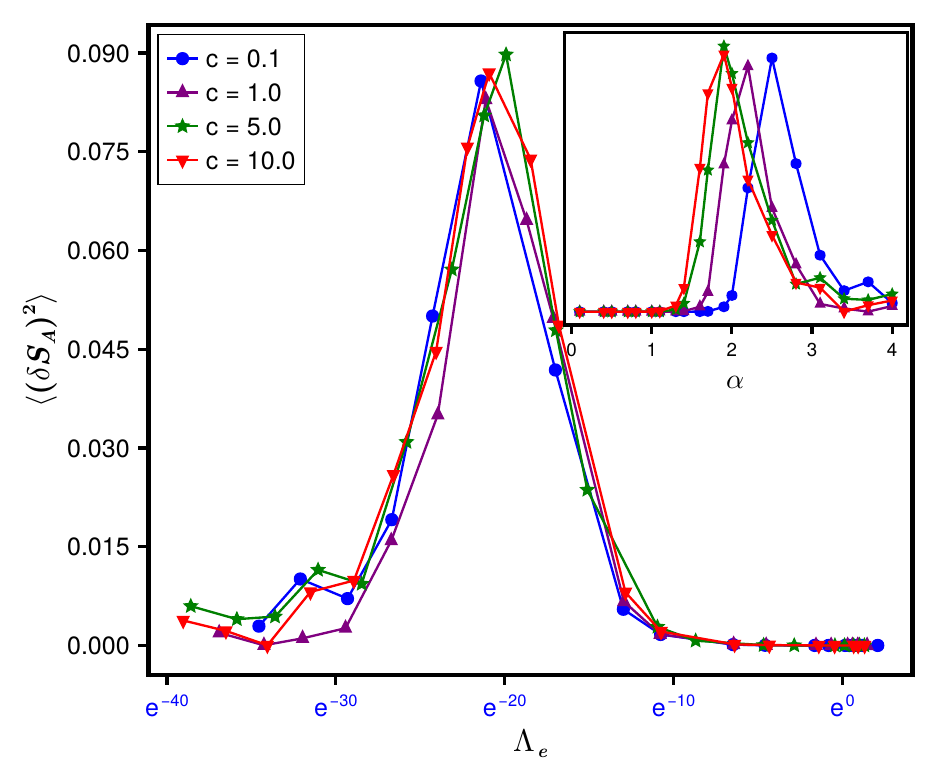}    
    \caption{\justifying \textbf{Evolution of SPEE measure for RP ensemble.} The evolution of average and variance of SPEE for the RP ensemble (with $c$ and $\alpha$ as the free parameters) with corresponding $\Lambda$ is shown. For comparison, the inset displays the evolution for different $c$ and $\alpha$ combinations, such that $\mu = c \, N^{\alpha}$ with $N = 12^3$.}
    \label{avg_r1_rp}
\end{figure}

The average SPEE with $\Lambda$ is shown in Fig. \ref{avg_r1_rp}, and the inset shows the corresponding behaviour with $\alpha$ for different $c$. As evident from the figure, the dynamics of average SPEE follows different evolutions curves for different $c$. Nonetheless, with $\Lambda$ they follow the same path. Indeed, the evolution of SPEE here shows a rapid transition from separable to maximum limit with $\Lambda$. 

\begin{figure}[h]
    \centering
 \includegraphics[width=0.7\textwidth]{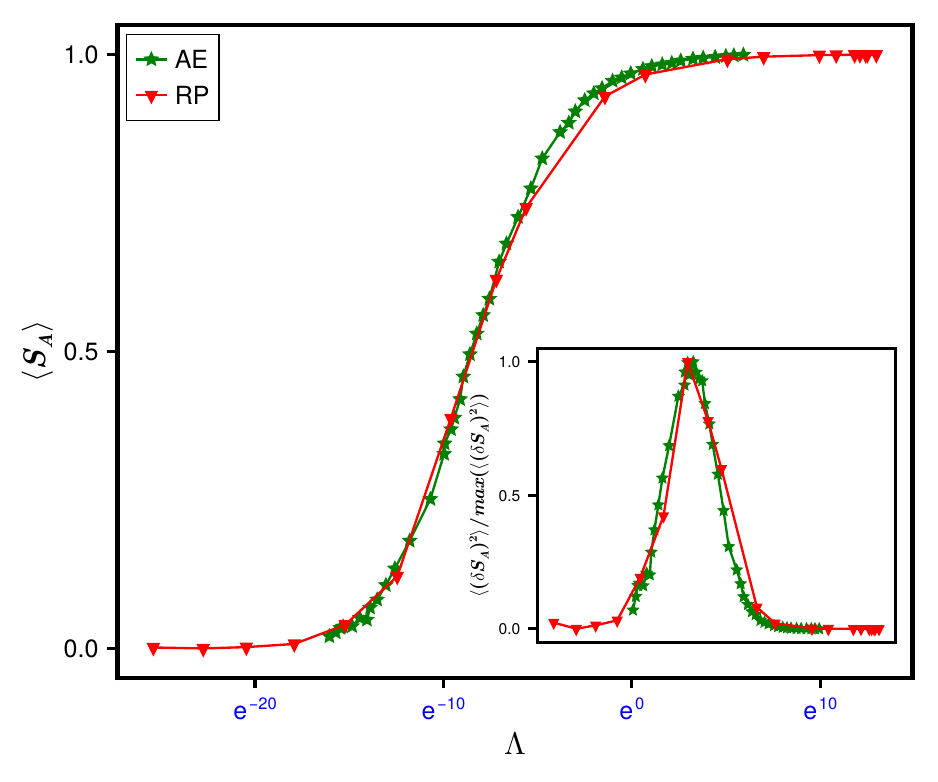}
    \caption{\justifying \textbf{Comparison of SPEE measures evolution for RP and Anderson ensemble.} The comparison of the evolution of average SPEE for one of the cases for the RP and Anderson ensemble (AE) is shown. The inset shows a similar comparison for the variance of the SPEE.}
    \label{sa-rp-and}
\end{figure}

Based on complexity parameter formulation, the analogy of $\Lambda$-governed evolution of SPEE measures is predicted not only for a given Hamiltonian under different system conditions, it is also expected to occur for two different Hamiltonians too as long as they belong to same class of global constraints e.g. symmetries and conservation laws (\cite{pseig}). As clear from figure \ref{rr1_ae} and figure \ref{avg_r1_rp}, the $\Lambda$ parameters for the two cases are different by a factor of $10^5$; a rescaling of $\Lambda$ for the RP case by $10^5$ is therefore needed for their comparison. Figure \ref{sa-rp-and} displays a comparison of $\Lambda$-governed  $\bk{S_A}$ behavior for RP ensemble and Anderson ensemble; the good agreement of the two curves once again confirms the $\Lambda$ governed analogy.

\section{Conclusion} \label{conclusion}

 In this work, we have analyzed the SPEE dynamics   in disordered Hamiltonians as the system conditions vary. The theoretical approach presented here is based  on  the complexity parameter formulation of the eigenfunction statistics  of the Hamiltonians represented by multiparametric Gaussian ensembles; the formulation was developed and investigated in detail in  \cite{pseig}. Our theory predicts the existence of a common mathematical formulation of the SPEE for a wide range of single-particle Hamiltonians  where system dependence appears collectively through the complexity parameter  $\Lambda$, a functional of the system parameters,  In particular, we have tested out theoretical prediction for many system conditions of two prototypical Hamiltonians, viz., the 3D Anderson model, and the Rosenzweig-Porter ensemble. 
 

A common mathematical formulation for the entanglement dynamics in single-particle systems has by itself many implications. The first and foremost among them is the existence of an infinite range of universality classes of the SPEE statistics governed by the complexity parameter. The universality of the formulation also reveals the lack of sensitivity of the entanglement to specific system details; it depends only on their collective information. An extension of this formulation for  multi-particle fermionic systems is also desirable as well as important. In this case, the preferred basis for writing the eigenfunction will be the Slater determinant. The entanglement can be then studied in the traditional sense where a bi-partition consists of a fraction of total number of particles. An interesting path to pursue will be to develop a single-parametric formulation of the entanglement dynamics for this case and compare with the one developed for the multi-particle spin systems.

\section{Acknowledgment}
We acknowledge National Super computing Mission (NSM) for providing computing resources of `PARAM Shakti' at the IIT Kharagpur, which is implemented by C-DAC and supported by the Ministry of Electronics and Information Technology (MeitY) and Department of Science and Technology (DST), Government of India.  One of the authors (P.S.) is also grateful to SERB, DST, India for the financial support provided for the  research under Matrics grant scheme and also acknowledge support from the ICTP through the Associates Programme (2020-2025). D.S. acknowledges financial support from the MHRD through the PMRF scheme.


\bibliographystyle{ieeetr}
\bibliography{references}

\pagebreak

\appendix

\end{document}